\documentstyle[preprint,aps,amssymb,epsf]{revtex}
\tightenlines 
\renewcommand{\ni}{\noindent} 
 
\newcommand{\beq}{\begin{equation}} 
\newcommand{\beqa}{\begin{equationarray}} 
\newcommand{\eeq}{\end{equation}} 
\newcommand{\eeqa}{\end{equationarray}} 
\newcommand{\dmu}{\delta\mu^{\star}} 
\begin{document} 
 \title{Simple Model of  Capillary Condensation in porous media.}
\author{S. M. Gatica$^1$, M. M. Calbi $^2$ and M. W. Cole$^2$}
\address{$^1$Departamento de F\'{\i}sica, Facultad de Ciencias Exactas\\
y Naturales, Universidad de Buenos Aires, 1428 Buenos Aires\\
Argentina\\
$^2$Department of Physics, The Pennsylvania State University,\\
 University Park, Pennsylvania 16802}

\date{\today}
\maketitle{}
\begin{abstract}

We employ a simple model to describe the phase behavior of $^4$He and Ar in a hypothetical porous material consisting of a regular array of  infinitely long, solid, parallel cylinders. 
We find that high porosity geometries exhibit two  transitions: from
vapor to film and from film to capillary condensed liquid. At low porosity, the film is replaced by a ``necking'' configuration, and for a range of intermediate porosity  there are  three transitions:
 from vapor to film, from film to necking and from necking to a capillary condensed phase.

\end{abstract}

\vspace{1cm}  
 
\section{
 Introduction}

The physics of adsorption  in porous media has long been appreciated for its  considerable importance in diverse applications, including gas storage, separations and cryopumping. In
addition, the geometry provokes intriguing fundamental questions about the
properties of phases of the adsorbate, for which the surface curvature energy
plays a prominent role \cite{Gubbins-review,chan1,chan2,tho,cs,diet,ev}. An ongoing question
pertains to the validity of numerous simple quasi-two or one dimensional models
which omit the heterogeneity present in  most such media. \cite{monson,miyahara} Moreover, there
remains a fundamental question about the nature and existence of genuine phase
transitions in such disordered media.

In this paper, we attempt to answer {\it qualitative} questions about the nature of such adsorption by employing a simple model of the geometry that we study with a
simple calculational procedure. The geometry is shown in figure 1 and described
in detail below. The procedure is called a ``simple model'' in a number of papers
published over the last decade \cite{slab,capillary,ancilotto,bojan,mistura,ancilotto2,cheng,curtarolo}.
In the model, the system's energy is taken as a straightforward sum of bulk energy,
surface energy and gas-solid interaction energy terms. The model is certainly
oversimplified; yet it has provided answers to questions about similar
adsorption problems which agree rather well with results obtained in more
reliable (even ``exact'') studies of the same problem. The rationale for applying
the present approach is that the results may yield broad trends which are
robust, at least qualitatively. In the present case, our results are phase
diagrams for which a key parameter is the porosity $\Phi$ of the system. As
expected, the physics of  pore filling involves a competition between adhesive
and cohesive forces. As such, the results of our calculations are sensitive to
assumptions we make about the gas-surface interaction. In the present case,
both these assumptions and the calculational results are analogous to those found in the
problem of wetting transitions. \cite{ross-bonn-review}

\section{The model}

We propose a simple model consisting of infinitely 
 long, solid, parallel 
cylinders of radius $R=30\AA$.
 The centers of the cylinders form a square lattice  
and the distance between sites, $S$, is related to  $\Phi$ by the equation

\beq
\Phi=1-\frac{\pi R^2}{S^2}.
\eeq 

\begin{figure}[ht]
\caption{Schematic diagram of the configuration; the unit cell is shown with dashed lines.}
\centerline{\epsfysize=8.in \epsfbox{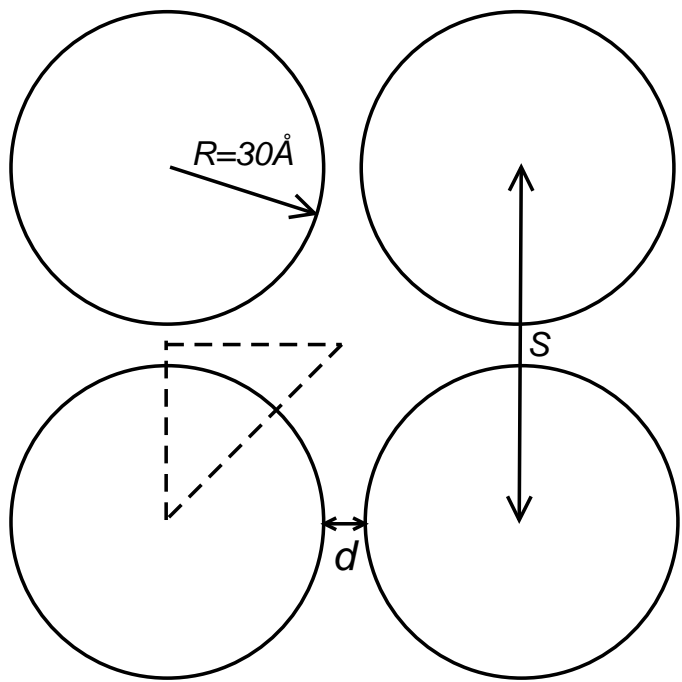}}
\label{configuration}
\end{figure}

\begin{figure}[ht]
\caption{a) Equipotential curves  due to the four closest  cylinders (with $\Phi=0.42$), for He. Contours are labelled by potential energy values, in meV. The dashed line shows a boundary of  the  unit cell.  
b) Potential contributed by a single  cylinder for He (dashed) and Ar (solid).}
\centerline{\epsfysize=8.in \epsfbox{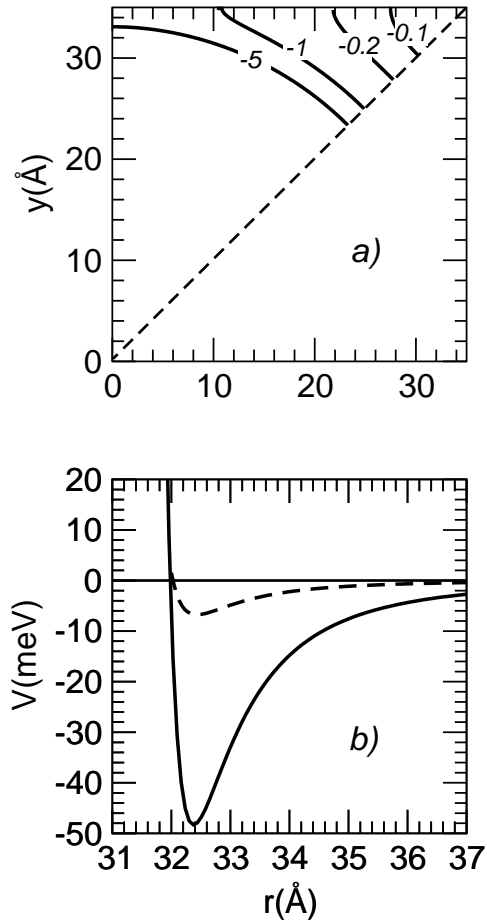}}
\label{potential}
\end{figure}

\noindent  The configuration is depicted in Fig. 
\ref{configuration}, where is also shown the triangular  
 unit cell considered in the calculations. The  interaction between the adsorbed  atoms  and the substrate  is approximated by the sum of the potentials from
 the four  cylinders that are closest to the unit cell. The potential due to each single cylinder, $V$, was constructed   assuming cylindrical symmetry  
so that it is a function of the distance to the axis only ($r$).   
The following function $V(r)$  reproduces both the potential of a flat substrate (which has often been  modeled as 
$4C_3^3/(27D^2 z^9)-C_3/z^3$, $z$ being the distance from the surface)
for $r\rightarrow R$ and the potential of an infinite wire for $r\gg R$,  
 
\beq 
V(r)=\frac{4C_3^3}{27D^2}\frac{1}{(r-R)^9}- 
\frac{C}{(C/C_3-R^2+r^2)(r-R)^3}, 
\eeq 
where  
$D$ and $C_3$ are the well depth and van der Waals parameters respectively \cite{chizmeshya,vidali} and  
$C=9\frac{\pi}{4}C_3 R^2$.  
We adopted values displayed in table \ref{tabla} that are intermediate between the strongly attractive graphite and the weakly attractive  alkali metals.
The form of $V(r)$                    
                       and the total interaction considering four cylinders are plotted in Fig. \ref{potential}.

\begin{figure}[ht]
\caption{Schematic  views of the film phase (left) and necking phase (right).}
\vspace{1cm}
\centerline{\epsfysize=2.in \epsfbox{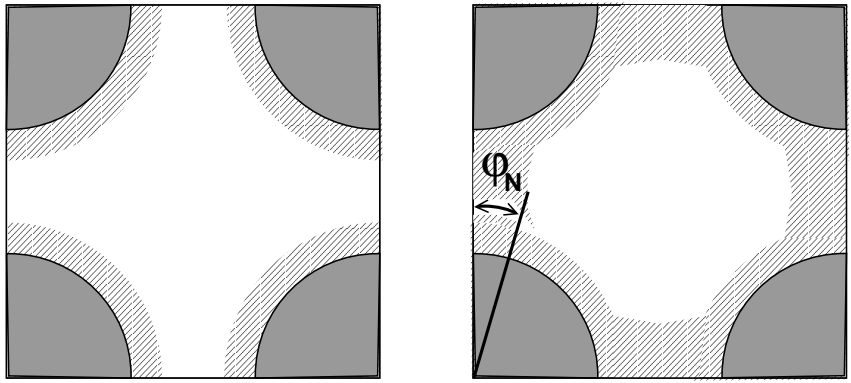}}
\label{phases}
\end{figure}

Atoms adsorbed in this material below saturated vapor pressure 
 are expected to form either a film adsorbed on the  
wall of the substrate,  or a condensed phase (C) filling all the free  space.  
Which of the two phases is stable can be determined by  evaluating and comparing the free energies. If the film is stable, and the porosity is moderately low ($<$50\%), the fluid may form bridges or ``necks'' between neighboring cylinders (see Fig. \ref{phases}). We will refer to this configuration as ``necking'' (N), and apply the term ``film'' (F)  
to the case where no necks are formed.  
The letter E refers to the empty pore  configuration. This actually means the presence of a low density vapor phase, which is ``empty''  only by  contrast to the higher density liquid phase.

For a given value of the chemical potential  $\mu$ below saturation ($\mu_0$),  
we evaluate the grand free energy per unit length  
 
\beq 
\Omega=F/L-\mu N/L. 
\eeq

\ni $F$ is the Helmholtz free energy, consisting of  
the substrate-fluid interaction, a bulk free energy and the surface energy.
 Assuming translational symmetry along the coordinate parallel to the axis of the cylinders, $z$,  the grand free   
energy per unit length of the fluid contained in the unit cell is,  
for the phase C 
 
\beq 
\Omega_{C}= 
 n\int_{0}^{\pi/4} d\phi \int_{r_{min}}^{S/(2\cos(\phi))} V(r)rdr- 
n(\mu-\mu_0)(S^2-\pi r_{min}^2)/8 +\sigma\frac{\pi}{4}r_{min}.
\eeq 
Here $\sigma$ and $n$ are   the bulk surface tension and number density of the fluid at a given temperature, and   $r_{min}$ is a cutoff distance, chosen as  the radius where  the potential due to  a single cylinder is minimum. We are taking the surface tension for the solid-liquid interface to consist of the sum of the  liquid-vapor interfacial tension and the integrated solid-fluid potential energy.
In this model, we are assuming  that a fluid of  density $n$  fills  all the available  space homogeneously , except the region  $r<r_{min}$.
  For  the   phase F we consider a homogeneous film of density $n$ formed between $r_{min}$ and $r_e(\phi)$, the energy  reads
 

\beq 
\begin{array}{ll} 
\Omega_{F}= & 
n \int_{0}^{\pi/4} d\phi \int_{r_{min}}^{r_e(\phi)} V(r)rdr+  
\sigma\int_{\phi_N}^{\pi/4}d\phi \sqrt{ 
r_e(\phi)^2+ r_e'(\phi)^2} -\\ 
&\frac{1}{2}n (\mu-\mu_0)\int_{0}^{\pi/4} d\phi (r_e(\phi)^2-r_{min}^2)
+\sigma\frac{\pi}{4}r_{min},
\end{array} 
\eeq 
 
\ni   where $r_e(\phi)$ is the equilibrium profile, and  $\phi_N$ is the angle subtended by the neck (see Fig. \ref{phases}) if present, taken to be  zero for a film. A prime in this expression means a derivative with respect to $\phi$.

By minimizing the energy $\Omega_{F}$, we get a differential equation for $r_e(\phi)$:  
 
\beq 
\mu-\mu_0-V[r_e(\phi)]=\frac{\sigma}{n  R[r_e(\phi)]}, 
\label{req} 
\eeq 
\ni where $R[r_e(\phi)]$ is the radius of curvature,  
 
\beq 
R[r(\phi)]=\frac{(r(\phi)^2+r'(\phi)^2)^{3/2}}{2r'(\phi)^2+r(\phi)(r(\phi)- 
r''(\phi))} 
\eeq

Notice that if  the surface tension  is small  the right hand side   
of eq. \ref{req} can be neglected and the following condition results  
 
\beq 
V[r_e(\phi)]=\mu-\mu_0. 
\label{equip}
\eeq 
This is the equation for   the  equipotential curves; its use corresponds to the venerable Polanyi theory of adsorption \cite{polanyi}.    
In the present study, we  use this relation, instead of solving eq.\ref{req},
 because it    greatly 
simplifies  the calculation. 
The approximation is good in a range of chemical potential such that 

\beq
\mu_0-\mu \gg\mid\sigma/nR[r_e(\phi)]\mid
\label{cond} 
\eeq
for all $\phi$. 
It will be shown below that in the case of He, the relevant values of $\mu$, where we find phase transitions,  fulfill this condition. For Ar, that is not the case at temperatures close to the triple point (i.e., when the surface tension is maximum).
These behaviors will be discussed 
in the next section.

\section{Results}

\begin{figure}[ht]
\caption{a) Grand free energies 
 for Ar at $T=130K$ ($\Phi=0.42$). b) 
 Number of atoms per unit length (solid line) and $dN/d\mu$ (dashed line,  in arbitrary units). The vertical line at the isotherm step corresponds to a delta function in this derivative.}
\centerline{\epsfysize=8.in \epsfbox{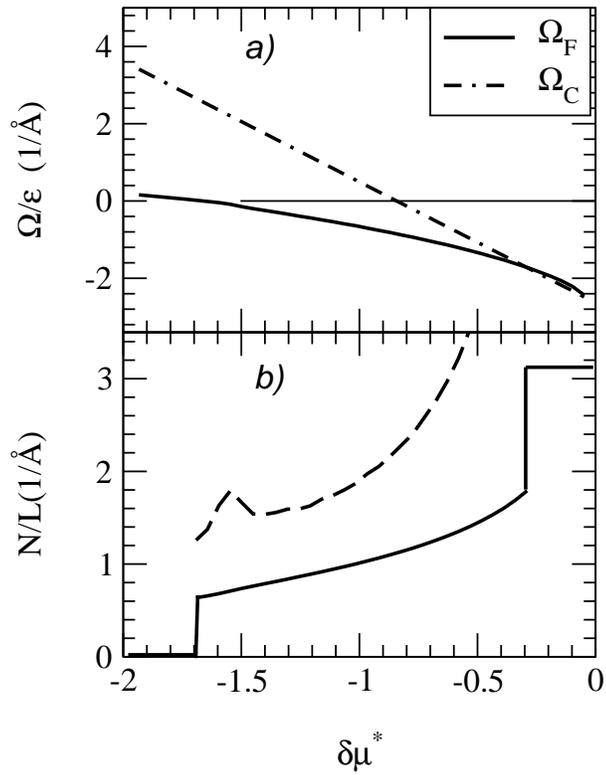}}
\label{enerar}
\end{figure}

\begin{figure}[ht]
\caption{Phase Diagram for Ar at $T=130K$. The dotted portion of the curves are explained in the text.}
\centerline{\epsfysize=8.in \epsfbox{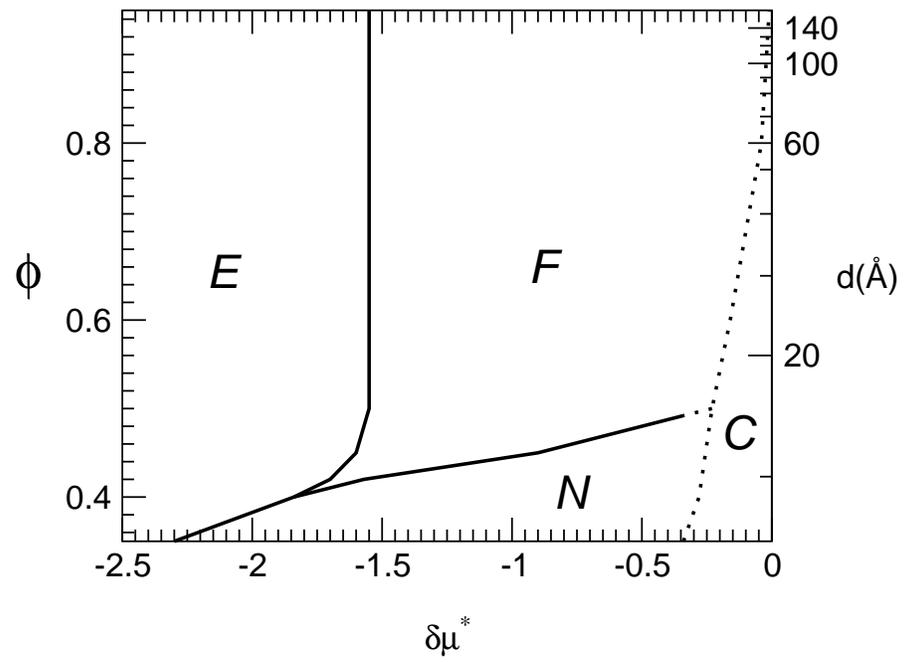}}
\label{pdar}
\end{figure}

In Fig. \ref{enerar} a) we display the grand free energy  of the phases F and C for Ar at $T=130K$ and porosity $\Phi=0.42$, which corresponds to a separation $d=10\AA$  between the cylinders.
Observe that   
 $\Omega_C$ and $\Omega_F>0$ for $\delta\mu^{\star} =(\mu-\mu_0)/\epsilon<-1.7$; hence the empty phase is stable in this  regime. However $\Omega_F<\Omega_C$ and $\Omega_F<0$  for  $-1.7<\dmu<-0.28$,  so the film is stable there, while for  $\dmu>-0.28$ the  completely filled phase is stable.  In Fig. \ref{enerar} 
b) we plot the corresponding isotherm, together with the derivative $dN/d\mu$. The steps in the 
isotherm correspond to the  transitions E$\rightarrow$F and  N$\rightarrow$C, and the peak in $dN/d\mu$ corresponds to  the 
transition F$\rightarrow$N .  From similar  analysis  for different  porosities, we construct the phase diagram shown in Fig.  \ref{pdar}. The line of the phase diagram is dotted in the region where the condition given by eq. \ref{cond} fails, according to the following criterion:  in the case of films the quantity $\sigma/nR[r_e]$ will be smaller than $\sigma/nr_{min}\equiv -\dmu_{min}$, so we   consider that the  condition (\ref{cond}) is fulfilled if $\dmu<10 \,\dmu_{min}$.
In the case of necking, the curvature radius for $\phi\simeq\phi_N$ is not necessarily larger than  $r_{min}$ and therefore  $r_{eq}(\phi\sim\phi_N)$ is poorly determined. However, the error introduced to the energy by this difference in the profile is expected to be small. 
For temperatures closer to the triple point, the surface tension (and therefore $-\dmu_{min}$) is larger,  and the transitions occur at values of $\mu$ such that $\dmu>10 \,\dmu_{min}$, so  the analysis done  using eq. \ref{equip}  becomes meaningless for such low temperatures. 
As is seen in Fig. \ref{pdar}, the phase N is possible only for very low porosities (small $d$), where the volume-filling ``price'' of necking is smaller. The upper part of the diagram is  qualitatively equivalent to the one obtained for a slab with large separation between walls. (see  Fig.4 in ref. \cite{slab})

\begin{figure}[ht]
\caption{Pore-filling behavior for helium at T=0 and 3K. Left panels show the   grand free energies $\Omega_F$ (solid line) and $\Omega_C$ (dashed line) for He at the temperature indicated. Right panels show the energy contributions to $\Omega_F$, due to the surface  (dotted line), substrate (dashed line) and volume (dotted-dashed line).}
\centerline{\epsfysize=7.in \epsfbox{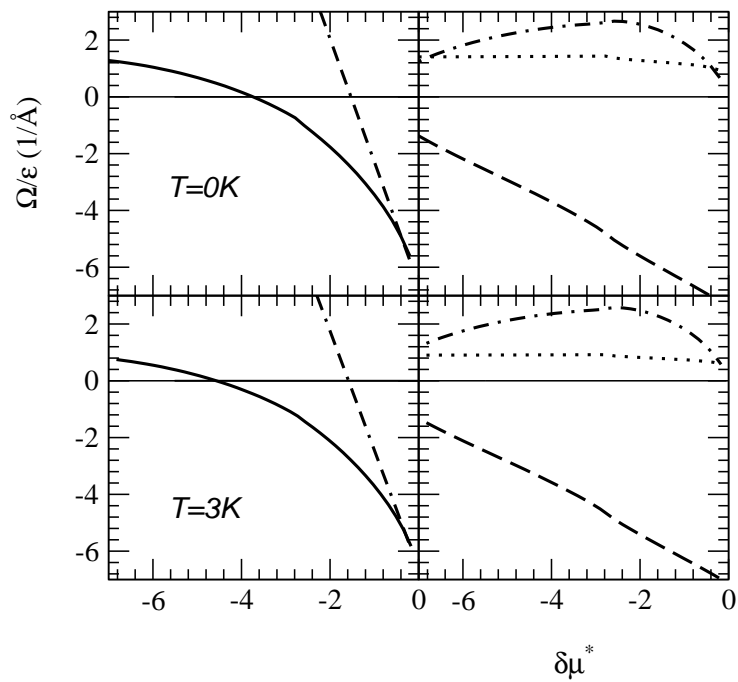}}
\label{enerhe}
\end{figure}

\begin{figure}[ht]
\caption{Phase Diagram for He, at the temperatures indicated. 
}
\centerline{\epsfysize=8.in \epsfbox{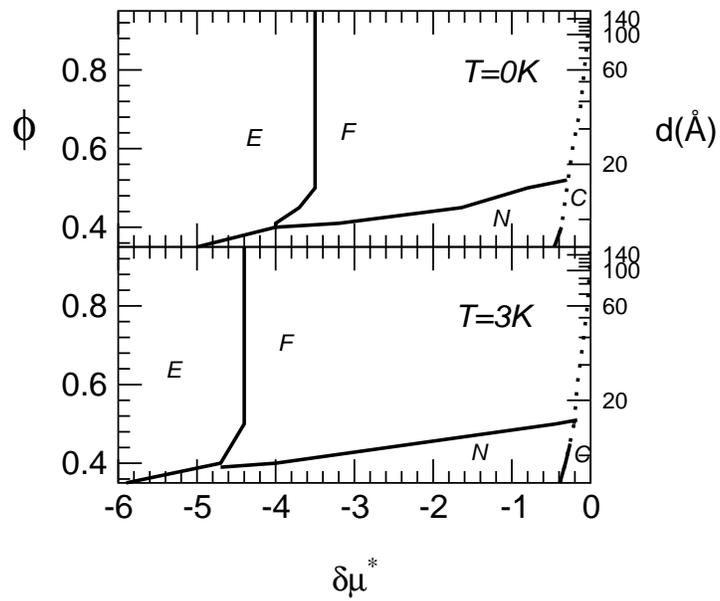}}
\label{pdhe}
\end{figure}

The corresponding  energies calculated  for He  are shown  in Fig. \ref{enerhe} for two different temperatures and $\Phi=0.42$, together with the components of $\Omega_F$, on the right. By means of the same procedure described above for Ar, we derived the phase diagrams displayed in fig.  \ref{pdhe}. The scenario is the same as for Ar. 
Comparing the results for $T=0$ K and 3 K, we  notice  that  at higher temperature there is a wider region where F is stable. 
This can be explained in terms of the values  of the  surface tension:  for T=3 K, $\sigma/n$ is $30\%$ smaller than for T=0 K. This is evident in Fig. \ref{enerhe}, on the right, where we can see that the contribution of the surface energy to $\Omega_F$ is more important for T=0 K than for 3 K. As a consequence, the phase F, that requires the formation of an interface, is more favorable  at higher temperatures. For Ar, the region of the phase diagram corresponding to the phase F is even narrower, as is expected since  the 
surface energy relative  to the substrate potential energy (in relative terms) is larger for Ar than for He at any temperature.

\section{Comments}

In this paper we have derived the phase behavior for gases exposed to an
array of parallel cylindrical strands of material. Using a number of
simplifying assumptions, this calculation yields a kind of generic
behavior, exemplified by the similarity of the behavior predicted for a
classical fluid (Ar) and an extreme quantum fluid ($^4$He). The behavior
is such that low porosity ($\Phi<0.4$) geometries exhibit a necking
transition at low chemical potential, followed by capillary
condensation. For a narrow range of intermediate porosity
($0.4<\Phi<0.5$), the behavior is different: a thin film forms, followed
by necking (with a singular derivative but no discontinuity in the isotherm), followed by
capillary condensation. At porosity $\Phi>0.5$, there is a transition from film to capillary condensation and necking does not occur 
 at any chemical potential. 
We note that the Kelvin equation does not agree 
with the present results for 
the threshold of capillary condensation. This discrepancy is well known from 
similar studies of small pores, due to the equation's neglect of the substrate potential 
and the resulting depleted density region near the wall (so that the pore 
radius is ill-defined).
\cite{evans}

This research has been funded in part by the Army Research Office, the Petroleum Research Fund, the American Chemical Society, CONICET  and the University of Buenos Aires.

\begin{table}
\caption{ Well depth and van der Waals coefficients $D$ and $C_3$ used in the calculation and parameters of the gas-gas interaction, $\epsilon $ and $\sigma $ from ref. \protect \cite{parameters}.}
\vspace{.5cm}
\begin{tabular}{ccccc}
  &$\epsilon(meV)$&$\sigma(\AA)$& $D(meV)$&$C_3(meV \AA^3)$ \\
\hline\\
Ar&10.34&3.40&50&1000\\
He&0.88&2.56&7&150\\
\label{tabla}
\end{tabular}
 \end{table}


\newpage

\newpage

\newpage

\newpage

\newpage

\newpage

\newpage



\end{document}